\documentclass[ 3p, onecolumn ]{elsarticle}
\usepackage{epsfig}
\usepackage{subfigure}
\usepackage{amssymb}
\usepackage{amsmath}
\usepackage{amsthm}
\usepackage{epstopdf}
\usepackage[table]{xcolor}
\usepackage[utf8]{inputenc}
\usepackage{graphicx}
\usepackage{placeins} 
\usepackage{outlines}
\usepackage{mathtools}
\usepackage{textcomp}
\usepackage{gensymb}
\usepackage{xcolor}
\usepackage{soul}

\begin{document}
\begin{frontmatter}

\title{The Effect of Blue and Infrared Laser Melting Frequency on Oxide Morphology in 304L}

\author[add4]{Jonathan S. Paras}
\cortext[corresponding]{Corresponding author}
\ead{jsparas@sandia.gov} 

\author[add4]{Randy D. Curry}

\author[add4]{James A. Ohlhausen}

\author[add4]{Michael J. Abere}

\affiliation[add4]{organization={Sandia National Laboratories},
            city={Albuquerque},
            postcode={87185-0959}, 
            state={NM},
            country={USA}}

\begin{abstract}

Continuous Wave Laser Beam (LB) melting offers control over the localized heating and cooling of melt pools for welding, brazing, additive manufacturing, and solidification. Research into laser-liquid metal interactions have primarily focused on the heat and mass transport under large thermal gradients imposed by the localized melting conditions. However, little research has been conducted into varying input laser frequency and scan rate at fixed absorption to understand the effects of laser-light on surface oxide formation. This article conducts laser melting of 304L under Blue (450 nm) and Infrared (IR, 1064 nm) laser frequencies and examines their impact on laser oxide thickness, chemistry, and coloration. We find that laser frequencies induce changes in the oxide layer thickness and chemistry that cannot be explained using conventional thermal absorption shifts and fluid dynamics. We suggest that light coupling may have thermodynamic implications for the chemical potential of the liquid metal which may drive  the observed phase behavior.

\end{abstract}

\end{frontmatter}

\section{Introduction}

Rapid heating and solidification under Laser Beam (LB) melting conditions has been the purview of laser welding, cutting, surface ablation, and more recently additive manufacturing of metallic materials\cite{schmidt2017laser,dupont2013welding,powell2012technical,fox2013novel}. Because of the range of cooling rates and spatially controlled energy input, LB melting serves as an attractive avenue for the design of metal manufacturing processes that can manipulate the microstructure, degree of atomic ordering, and non-equilibrium phase formation to produce high-performance metallic parts \cite{murr2012metal}. 

Scientific LB research in the US initially focused on using constitutive heat and mass transfer modeling to explain observed melt-pool geometry and chemical flow under temperature and composition gradients with the goal of predicting final microstructural evolution\cite{david2003welding,crossphysics}. Research was also conducted to couple the driving forces provided by changes in the chemical potential under such processing conditions to improve the accuracy of both the phase formation model and microstructural evolution\cite{elmer1991modeling,elmer1989microstructural}, largely because constitutive heat and mass transfer modeling alone failed to predict non-equilibrium structures that consistently formed in such melt pools. 

In parallel, research in the Soviet Union began to focus on the potential for optothermodynamic effects, that is, the coupling between the induced laser energy and the thermodynamic behavior of the melt. Research focused on high-powered (10$^{7}$ $\longrightarrow$ 10$^{8}$ W/cm$^{2}$), non-resonant regimes where differences in the absorption coefficient of liquid metals were not expected. Liquid metals and alloys were found to eventually undergo metal-to-dielectric transitions of the kind discussed by Hensel and Warren \cite{hensel2014fluid}. These transitions were thought to occur because significant laser irradiation near the Clausian liquid-vapor line resulted in such large electronic fluctuations that the electronic structure would come to resemble a dielectric, thereby becoming transparent to irradiation and preventing the infinite heating of liquid metals by laser irradiation alone\cite{batanov1973evaporation,prokhorov1973metal,bunkin1980nonresonant}. 

Previous work has studied the formation of oxides under arc-welding, native growth, and IR laser melting conditions in both inert and non-inert environments \cite{ling2019investigation,allen1988surface,goutier2011304l,adams2013nanosecond}. Oxide layer growth in these studies resulted in Cr-rich oxides predominating at the surface with thicknesses ranging from 20-500 nm.  To date, little work has been conducted to examine the effects of different continuous-wave (as opposed to nano or fhemtosecond pulse) laser frequencies, rather than intensity, on the resultant phase and chemistry of the melt pool and its oxide, beyond of course those that are phenomenological (additional heat input, melt-pool geometry, etc). In this article  we seek to conduct a simple experiment to examine how oxide chemistry and morphology changes for a plate of 304L under varying frequency irradiation and identical power. We hypothesize that the change in oxide chemistry and morphology will service as a quantifiable analogue to laser induced changes in the chemical potential of the liquid metal because the gas atmosphere, absorption spectrum, incident power, and base alloy chemistry will be similar under Blue and IR frequency laser irradiation. 

To interpret our results, in the absence of additional experimental data on both the absorption spectrum of liquid metal alloys, electronic transport properties in their liquid, and thermodynamic data that might also substantiate some degree of ordering, we will carry forward the assumptions made by the preceding literature not out of endorsement, but rather out of necessity. While they may temper the certainty of our conclusions, a clear avenue for necessary future work is thus illuminated.

\section{Methods}

Laser irradiation was performed within a laser engineered net shaping additive manufacturing (AM) tool. The tool had two optical ports on which a 2 kW YLS-2000 laser operating at a wavelength of 1064 nm (IPG, Photonics, Oxford, MA) and a 600 W Nuburu fiber laser operating at 450 nm was mounted to the spindle of a 3-axis Tormach CNC 770 (Tormach LLC., Waunakee, WI). The stages were housed in a controlled atmosphere glovebox (MBruan, Inc., Strathan, NH) with a continuously purging Ar gas feed to maintain an atmosphere of $<$50 ppm O$_{2}$ and $<$10 ppm H$_{2}$O. Further information on the tool can be found in Ref. \cite{kustas2018characterization}.  While the AM tool is capable of flowing powder, laser irradiation of the 1 mm thick 304L stainless steel sheet (Goodfellow) cut into 1 cm$^{2}$ squares was performed with the powder flows inactive. Samples of 304L were irradiated in a serpentine pattern over a 1 cm$^{2}$ area with a background preheat temperature of 385 °C. Both spatial Gaussian lasers were focused through 10 cm plano-convex lenses to a 1/e$^{2}$ radius of 0.6 mm. Beam spot sizes were measured using a WinCamD beam profiling camera. Hatch spacing between each laser scan was held constant at 0.5 mm. Laser parameters were chosen to match the solid absorbed intensity between the two lasers\cite{jyothi2017optical} to match the typical assumption in selective laser melting models of constant absorptivity \cite{khairallah2016laser}. A test matrix of processing parameters is given in Table (\ref{TestMatrix}).

\begin{table}[]
\centering
\label{TestMatrix}
\begin{tabular}{ccccc}
 \\
Laser Type & Hatch Spacing   &  Output Power  &  Traversal Rate &  \\
& (mm) & (W) & (mm/s) & \\
\hline
Blue & 0.5  & 100 & 1  &  \\
Blue & 0.5  & 100  & 2  & \\
Blue & 0.5  & 125  & 5 &  \\
IR & 0.5 & 116  & 1  &  \\
IR & 0.5  & 116  & 2 &  \\
IR & 0.5  & 145 & 5 \\
\end{tabular}
\caption{A test matrix of melting parameters for 304L experiments}
\end{table}

\begin{figure}
    \centering
    \includegraphics[width=1\textwidth, height=.75\textheight]{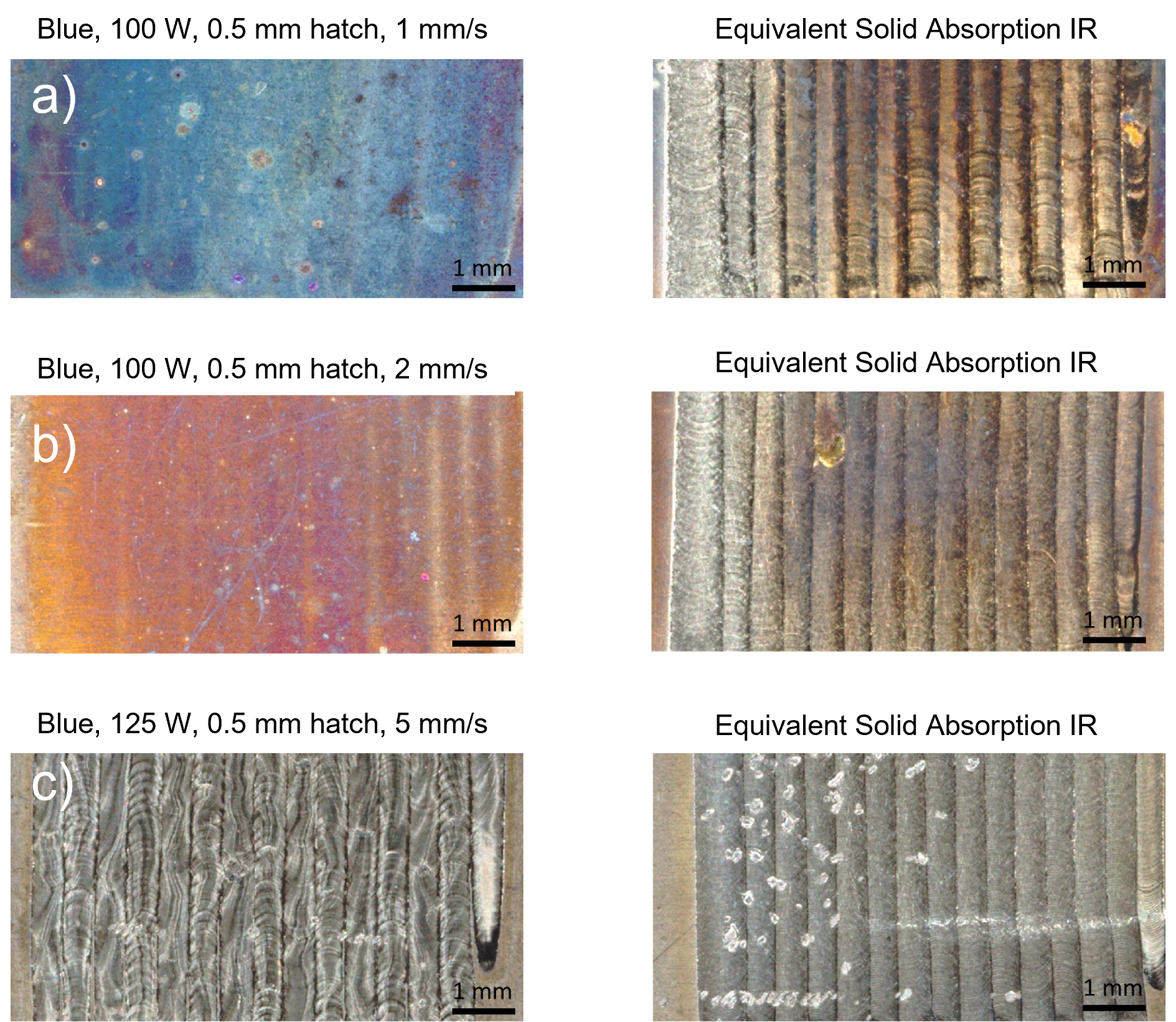}
    \caption{Optical characterization of the melt-track studies under the conditions outlined in Table (\ref{TestMatrix}). Note the optical contrast between Blue and IR laser melted samples, indicative of characteristic differences in the laser oxide layer chemistry and thickness}
   \label{OpticalChar}
\end{figure}

Characterization of the samples involved the use of Time-of-Flight Secondary Ion Mass Spectrometry (ToF-SIMS) to study the chemistry and its spatial distribution within the oxide layer. Analysis was conducted in an TOF-SIMS.5-100 instrument by IONTOF GmbH (www.iontof.com) with a base pressure of $<$ 1 x 10$^{-9}$ Torr. Depth profiles were performed in the center of each sample’s laser melted region by alternating 25kV, 1.5pA, high-current bunched mode Bi1+ analytical ions rastered over 50x50 µm$^{2}$ and 1kV, 45nA, Cs+ sputtering ions rastered over 400x400µm2.  Negative secondary ions were collected in this experiment in order to differentiate between metal oxide and bulk chemistries. This technique can measure multiple layer thicknesses as well as chemical gradation throughout the oxide layer as a function of thickness. Further information on the Tof-SIMS tool and method is provided in Refs \cite{schindelholz2018electrochemical} and \cite{ohlhausen2013tof}.

\section{Results}

A single pass melt threshold study was conducted to verify the power conditions under which melting will occur for 304L under 450 nm and 1060 nm laser irradiation. Power was swept from 25 to 200 W in increments of 25 W . The scan speeds were swept from 5-8 mm/s in increments of 1 mm/s. Multi-pass experiments were also performed within a power range of 25 W to 125 W with a 0.5 mm hatch spacing at 1-6 mm/s scan speeds in increments of 1 mm/s. Using a Keyence VHX-6000 digital microscope and a VK-X3100 Laser scanning microscope, melting was verified to occur at 75 W at 8 mm/s on a 1 in$^{2}$, 1 mm thick sample. Given that the addition of serpentine scan patterns adds incubation effects upon subsequent passes, the minimum laser dose for melt will not decrease. Thus, characterization of oxides arising from laser interactions with liquid metal focused on irradiation with at least 100 W of blue light with a scan velocity below 6 mm/s. Optical images from the serpentine scanning experiments at each of the dose conditions outlined in Table (\ref{TestMatrix}) are shown in Figure (\ref{OpticalChar}). 

Laser induced oxide growth is compared between the two lasers at 100 W at 1 mm/s and 2 mm/s in Figure (\ref{OpticalChar}) a and b, respectively. The self-consistent beam profiled spot size ensured that radial thermal conduction away from the spot center remained consistent for comparative analysis. The factor of two in total dose changed the oxide grown from blue laser irradiation from a blue to an orange-brown layer that was opaque enough to occlude the underlying scan tracks. For an equivalent solid absorbed power with constant hatch spacing and scan velocity, the oxide appears optically brown while the laser scan tracks remain visible beneath it. A third irradiation condition at 125 W at 5 mm/s (another factor of 2 decrease in total dose) was also characterized for oxide growth if Figure (\ref{OpticalChar})c because it represented a case in which the blue laser grown oxide does not occlude the underlying scan tracks. Where visible, there are not mesoscopically visible boiling voids on any of the track surfaces similar to those imaged optically in Ref. \cite{shrestha2019study}. 

\begin{figure}
    \centering
    \includegraphics[ height=.75\textheight, width=1\textwidth]{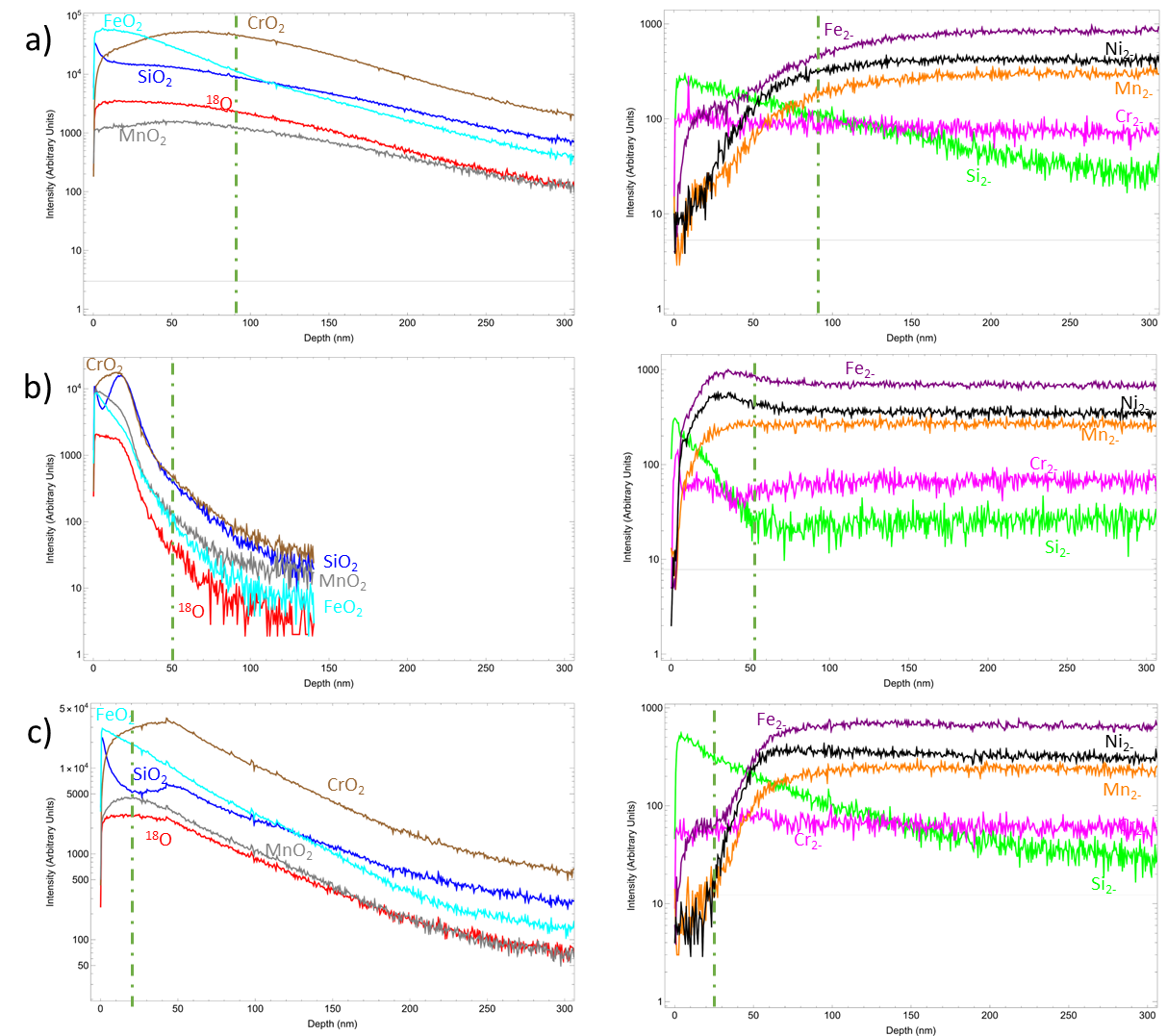}
    \caption{Time of Flight SIMS depth profile data for a) Blue, 0.5, 100 1 mm/s, b) Blue, 0.5, 100, 2 mm/s, c) Blue, 0.5, 125, 5 mm/s samples. Vertical dashed lines indicate the identification of a notional oxide layer. The left plot shows the depth dependence of the oxide components and the right plot indicates metal ion chemistry depth dependence.}
    \label{TofSims}
\end{figure}

\begin{figure}
    \centering
    \includegraphics[ height=.75\textheight,width=1\textwidth]{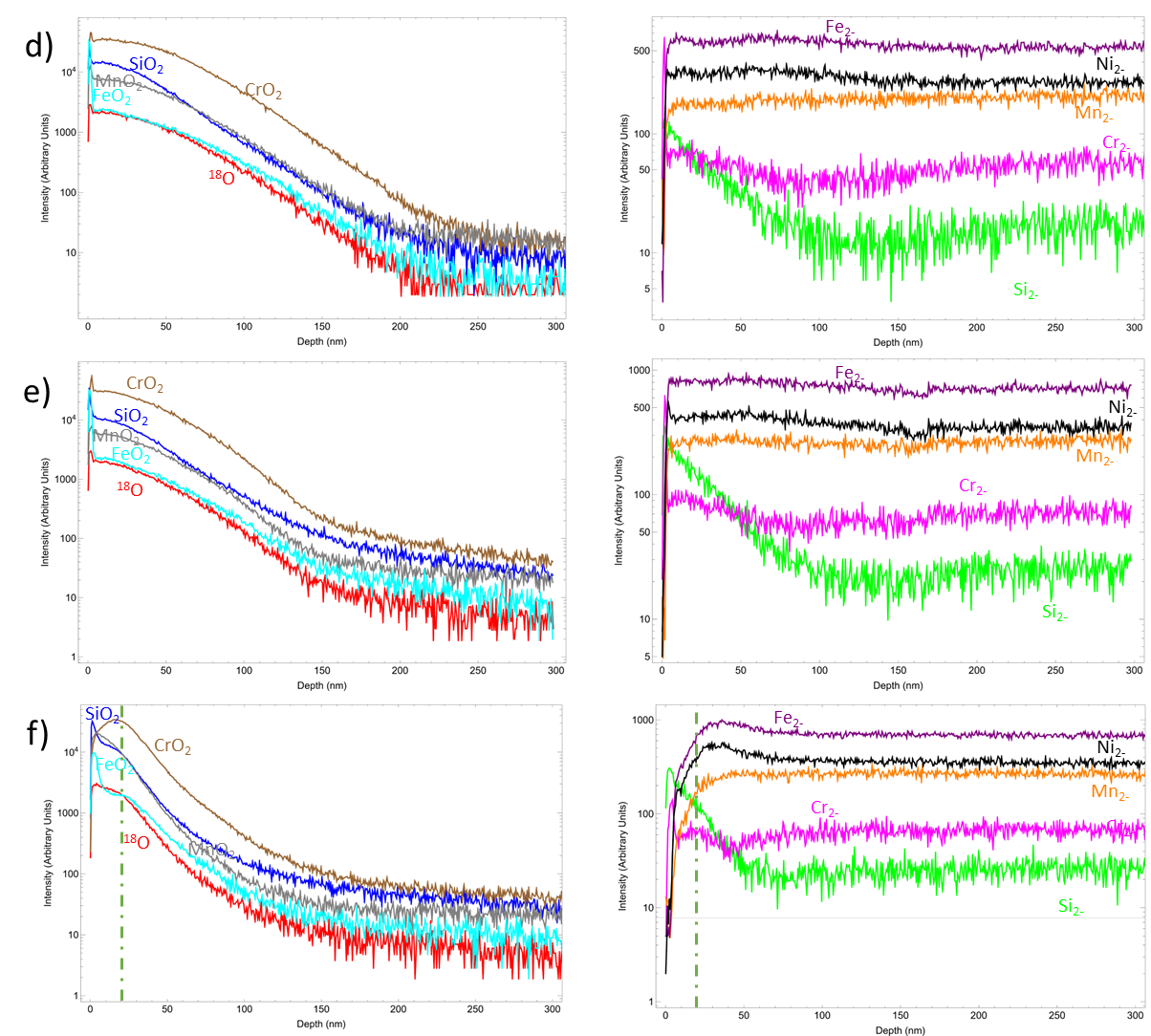}
    \caption{Time of Flight SIMS depth profile data for d) IR, 0.5, 116w 1 mm/s, e) IR, 0.5, 116 W, 2 mm/s, f) IR, 0.5, 145 W, 5mm/s samples. Vertical dashed lines indicate the identification of a notional oxide layer. The left plot shows the depth dependence of the oxide components and the right plot indicates metal ion chemistry depth dependence. For the IR samples, the oxide layer was either very small ((f), 20 nm) or difficult to distinguish from an oxygen rich mixed metal layer (d, e). }
    \label{TofSims2}
\end{figure}

Characterization of the test matrix in Table (\ref{TestMatrix}) of oxidized 304L plates was conducted via ToF-SIMS. Spectra were taken from the center of each coupon to measure a condition with self-consistent incubation effects. The results are reported herein. Blue laser (450 nm) results indicate a notional oxide layer that is qualitatively different in structure, chemistry and thickness compared to IR (1064 nm). The blue laser produced thicker oxide layers whereas IR irradiation produced such oxygen deficient layers that were so thoroughly mixed with the substrate chemistry that a notional oxide layer could not be discerned except in the case of samples with the fastest laser track rates (5 mm/s). The oxides were relatively iron rich under Blue irradiation, compared to increased Cr$_{2}$O$_{3}$ in IR. The IR laser produced oxide layers that were difficult to distinguish with the base metal and therefore oxygen poor. ToF Sims data is provided in Figure (\ref{TofSims} and \ref{TofSims2}). Our data consistently indicates that increased heat flow generally results in oxides that are thinner. A schematic depiction of the oxide layer morphology and relative thickness is given in Figure (\ref{appendixCartoon}) in the Appendix.

\section{Discussion}

Differences in the morphology and thickness of the oxide layer produced at power conditions near 10$^{4}$ W/cm$^{2}$ cannot be explained using our conventional understanding of laser absorption spectra in liquid metals and optothermodyanmics under significant illumination (10$^{7}$ $\longrightarrow$ 10$^{8}$ W/cm$^{2}$) alone \cite{batanov1973evaporation}. We will examine each of these possibilities in depth first and then propose a different possible explanation for the observed changes under changing laser, which the experimental conditions indicate to be non-resonant, low-power conditions.  

\subsection{Absorption Coefficient}

Data for the absorption spectrum in 304L in the domain of interest is, to the best knowledge of the authors, nonexistent. We will therefore be forced to develop an approximation for the absorption spectrum to determine the significance of any difference in the optical absorption spectrum and its impact on the observed phase behavior.

The optical properties of materials can be calculated analytically if the complex and real components of the refractive index and by extension, the relative permittivity are known\cite{fox2010optical}. The Drude free-electron model of conductivity enables the analytical evaluation of the relative permittivity, and this has demonstrated some efficacy in describing the absorption spectrum in materials like Al and as well as the alkali and alkaline-earth metals. 

High-temperature research has suggested that such assumptions may apply to some liquid metals, such as Al, Li or K\cite{comins1972optical,miller1969optical}. However the work of Miller suggests that while such a model may be descriptive for some p-block and s-block metals, transition metals like Fe, for which the order-of-magnitude in absorption spectrum may be described by a Drude oscillator model, may not exhibit a monotonically decreasing frequency dependence after the plasma frequency \cite{miller1969optical}. While we will make this assumption here, this is out of necessity due to the lack of data of the high-temperature liquid properties of stainless steel at our specific test wavelengths, rather than out of endorsement for this assumption. We will discuss later the ramifications of ignoring such a possibility, tempering our proposed conclusions.

\begin{figure}
    \centering
    \includegraphics[width=\textwidth]{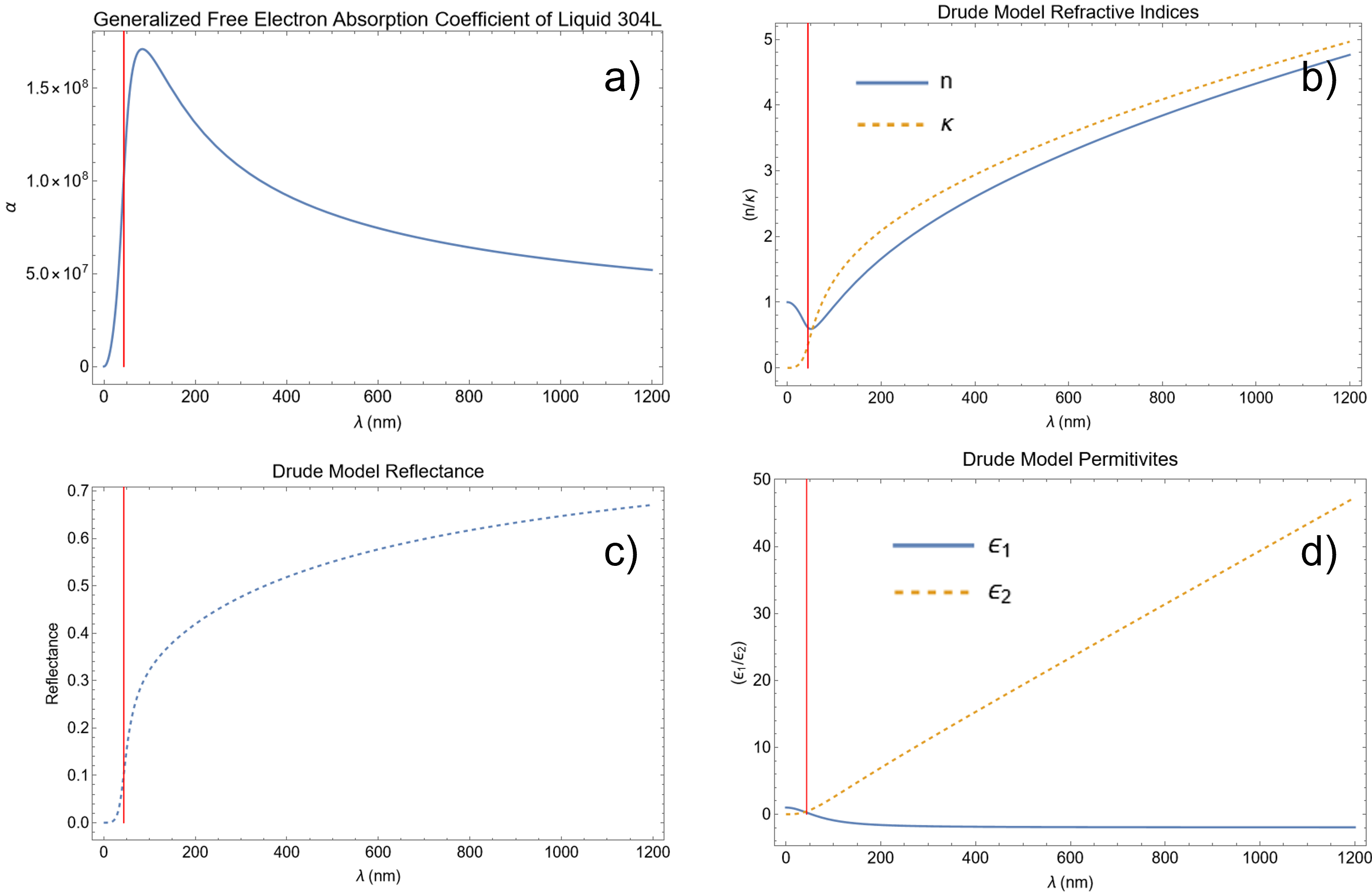}
    \caption{Absorption coefficient as a function of frequency for liquid 304L under Drude free-electron model of conductivity (a), the refractive indices (b), Reflectance (c), and the real and imaginary components of the relative permitivity (d). All valence electrons of Ni, Cr, Fe, and Mn were considered. The red-line indicates the plasma frequency, just beyond the main resonance expected in such an oscillator model for absorption, which occurs at roughly 43.5 nm.}
    \label{AbsorptionSpectrum304L}
\end{figure}

The spectrum presented in Figure (\ref{AbsorptionSpectrum304L}) shows that beyond the plasma frequency, liquid 304L is in a regime that should not experience a significant change in the absorbed energy. The absorption coefficients for 450 and 1064 nm from our model are 8.68 x 10$^{7}$ m$^{-1}$ and 5.53 x 10$^{7}$ m$^{-1}$ respectively. Our Reflectance values from Figure (\ref{AbsorptionSpectrum304L})c compare favorably to the literature at 630 nm (ours 0.58, literature 0.59), 600 nm (ours 0.578, literature, 0.62), and 1064 nm (ours 0.65, literature 0.69) \cite{kraus1989experimental,yang2021modeling}.

If we assume that the intensity $I$ falls off as Equation (\ref{AbsorptionEquation}), and assume a shallow melt depth from direct optical coupling of 100 nm, the absorbed intensity is 99.983$\%$ and 99.6$\%$. The assumed melt depth is quite conservative, and demonstrates that the energy deposition, on the length scale of the size of the oxide, is nearly complete. 

\begin{equation}
\label{AbsorptionEquation}
I = I_{0}e^{-x \alpha}
\end{equation}

The intensity is proportional to the absorbed heat. Laser wattage is proportional to the temperature increase and is given by the relationship in Equation (\ref{HeatCapT}).

\begin{equation}
\label{HeatCapT}
I \propto C_{p}\Delta T
\end{equation}

The experimentally measured heat capacity of liquid iron, as an analogue to 304L, is 825 J/kg K \cite{beutl1994thermophysical}. By assuming that the volume of the melted region is a cylinder with a radius of 0.5 mm and 100 nm depth, and also assuming that despite the high-scan speeds, roughly 100 J of energy for the Blue laser and 116 J of energy for the IR is deposited entirely in the melt, we find that the change in liquid temperature from the melting point would be 16 $\%$ higher in case of the IR.

We also examined differences in the absorption coefficient of the solid oxide and whether this may offer an explanation for the observed differences in oxide morphology due to differential heating based on solid oxide absorption spectra. The optical constants for \(\text{Fe}_2\text{O}_3\) are explicitly given in Querry where at 450 nm there is an absorbing oxide with \(R=0.264\) ($\mathcal{T}$ or transmission is \(10^{-11}\)) of 0.736 of upstream power \cite{querry1985optical}. This is consistent with published band gaps of \(\text{Fe}_2\text{O}_3\) of 1.9 to 2.2 eV in Sivula et al. and Bassi et al \cite{bassi2014iron,sivula2011solar}. The \(\text{Cr}_2\text{O}_3\) has a band gap of 3.1 eV from Abdullah \cite{abdullah2014structural}. So we would expect the blue light to selectively absorb in any \(\text{Fe}_2\text{O}_3\) that formed on the surface of the liquid as a slag. At 1064 nm, Querry places the reflectance ( of \(\text{Fe}_2\text{O}_3\) at 0.184 and a transmittance of 0.638. In the presence of any surface \(\text{Fe}_2\text{O}_3\), you would expect 0.178 to absorb directly in the oxide at the skin depth (\(\alpha = 4488.0 \, \text{cm}^{-1}\)) and then \(0.638 \times (1-0.65) = 0.223\) is absorbed. For oxide thickness less than the skin depth, the absorption in the underlying liquid will approach \((1-0.184) \times 0.35 = 0.286\) of incident power. It is a fair assumption that with thermal conductivity moving melt fronts 3 orders more than absorption depths that the bilayer of transmitting oxide plus surface melted zone would rapidly equilibrate in temperature. Due to the offsets in solid absorptivity, you end up with a case of 74 W vs 33 W of absorbed intensity with a nanoscale oxide with the Blue and IR laser respectively.

We then examined the implications for the oxide layer. For differential absorption to affect melt behavior, the oxide would have to form while the laser pulse was active on the liquid metal. We will examine what such differential heating, if absorbed by an oxide layer near the maximum thickness observed in the paper (100 nm), would change the differential heating of the sample based on the differential in power input for \(\text{Fe}_2\text{O}_3\) and \(\text{Cr}_2\text{O}_3\) using the equation
\begin{equation}
\Delta Q = \int_{RT}^{T_m} n_{\text{moles}} C_{p,\text{solid}} \, dT + n_{\text{moles}} \Delta H_{\text{fusion}} + \int_{T_m}^{T'} n_{\text{moles}} C_{p,\text{liquid}} \, dT
\end{equation}
In both cases, it was found that the oxide layer would increase in temperature 5 orders of magnitude beyond the melting point, implying that the small oxide layers would boil. Therefore, we find it unlikely that differential absorption of the \(\text{Fe}_2\text{O}_3\) for the blue laser light explains the observed phenomena. The oxide layers are thinner with the IR laser, despite the expectation that oxide stability should decrease with increasing temperature as indicated in an Ellingham diagram. Thinner oxide layers are therefore observed where we expect absorption to be great. It is still possible that a thin oxide layer may exist on the liquid metal, but this layer likely quickly thermalizes with the metallic bath below.

The final consideration is whether such temperature differences are sufficient to manipulate the thermodynamic equilibrium between the liquid metal and the observed oxide. Full thermodynamic modeling is necessary to quantify the quantity and chemistry of phases that emerge after LB melting. However, it is illustrative to examine just the effects on the partitioning of Fe between the liquid (or solidified) alloy and the oxide top-layer. This equilibrium is given by Equation (\ref{ChemicalPotentialFe}).

\begin{equation}
\label{ChemicalPotentialFe}
\mu_{Fe}^{o} + RT ln (a_{Fe,304L}) = \mu_{Fe}^{o} + RT ln (a_{Fe,oxide})
\end{equation}

If we assume that the solution thermodynamic behavior of both the liquid metal and oxides are ideal, under LB melt conditions presented herein, the difference in the increase in temperature would be tens of degrees, which is insufficient to influence the oxide chemistry, or to change the thermodynamic stability of one oxide phase against another. Furthermore, such self-similar melt temperatures would not change the Marangoni flow nor recoil pressure \cite{khairallah2016laser}, which eliminates changes in the fluid dynamics as the underlying mechanism for the observed differences in oxide growth.   

In the absence of any observed ordering in the melt that could provide associates capable of coupling to the laser light at higher frequencies, additional coupling of laser light at higher frequencies is an unlikely to cause a larger temperature difference between the Blue and IR laser frequencies. Conventional differences in the rate of oxide absorption, growth, and mass transport through both the liquid and solid cannot serve as an explanation for the observed differences in oxide morphology between the two frequencies. Additionally, the oxide layers under IR laser melting were found to be Cr-rich compared to relatively more Fe-rich layers under blue-laser illumination. The IR results seem to be consistent with oxide layers that form under both arc-welding conditions in air and nanosecond IR laser pulse studies, further suggesting that the Blue laser frequency may be altering the chemical potential sufficiently to change the oxide chemistry \cite{ling2019investigation,adams2013nanosecond}.

\subsection{Optothermodynamic Effects}

Bunkin has written a thorough review of optothermodynamic phenomena in liquid metals under laser irradiation \cite{bunkin1980nonresonant}. He proposed that under high-enough power , extreme excitation of the electrons in molten metal result in a phase transformation to a dielectric that is transparent to the laser light. This finding requires that the system does not hit the liquid-vapor equilibrium line or some type of liquid-liquid miscibility gap before a dielectric transition occurs. Such a concept is discussed in the monograph by Hensel in his monograph "Fluid Metals" and a more thorough review of the coupling of optical and thermodynamic properties can be found between both Hensel and Bunkin's treatments\cite{hensel2014fluid,bunkin1980nonresonant}. 

If a dielectric-to-metal transition were to occur in our system, that would also change the chemical potential of the constituent elements and could explain why oxide layer thickness and chemistry may change. But as has been stated elsewhere, we are operating at power-levels on the order of 10$^{4}$ W/cm$^{2}$, whereas Bunkin demonstrates such power-induced phase transitions occur near 10$^{7}$ W/cm$^{2}$. In our experiments, we do not reach conditions where a highly constrained liquid surrounded by solid material induces significant effective pressures on the molten region that would cause a phase transition. 

However, it is worth nothing that the work of Bunkin and Hensel demonstrates there exists a practical limit at which a laser can no longer be used to impart more energy into a liquid metal system with a classically thermodynamically defined upper limit near the liquid-vapor transition point. The data for such high-temperature phase boundaries in commercial alloy chemistry is poor, the liqudus lines are qualitative in nature and the phase boundaries of the liquid-vapor equilibrium are often lacking entirely \cite{hensel2014fluid}. The results presented herein motivate the instrumentation of high-temperature phase boundary and optical property measurements in liquid metals.

In the case of our experiments, we are three orders of magnitude below the limits defined for pure liquid iron by Batanov and Bunkin and do not expect this type of classical optothermodynamic effect to be driving the observed differences in oxide layer chemistry and morphology between the different laser frequencies \cite{batanov1973evaporation}.

\subsection{Electronic Structure Effects}

Previous work has demonstrated that for systems near dielectric-metal transitions, optical illumination can induce the phase transition in semiconductors, but to the best knowledge of the authors, such an observation has not been made in liquid metals under low-power conditions \cite{tao2016nature}.  

Our work demonstrates that the oxide layer chemistry can change significantly as a function of laser frequency, not by additional heat absorption or other resonant behavior, in a regime where classical defined vapor transformations and dielectric metal transitions need not apply.  There have also been experiments conducted that examined the metal-phase that emerges under different laser frequency irradiation in the Al-Cu system. It was found here that metallic phases -- described as a "highly strained structure" that did not index in electron backscatter diffraction and exhibited microhardness akin to a supersaturated Al in Cu phase, which suggests that these phases were non-equilibrium in nature \cite{abere2024refining}. 

Understanding the induced changes in the activity coefficient in the liquid metal phase in the absence of conventional explanations points to the possibility that the laser irradiation may be significantly altering the electronic states, and perhaps the entropy, directly. In Abere et al \cite{abere2024refining}, we have corroborating evidence from the solidified non-equilibrium structure  not observed at the Cu/Al interface in IR laser welds \cite{lee2014effect} \cite{zhang2020laser} that complements our findings for the oxide. Our explanation for the observed behavior introduces several important questions that will need to be answered before we may substantiate this theory. 

If we have been able to manipulate the distribution of accessible electronic states or entropy of the electron subsystem by changing the laser frequency, without significantly altering the total energy state of the system,  upon excitation, electrons then must be moving to a region of state-space where the density of states is identical. Otherwise, one would expect that electronic excitation should result in an increase in energy absorption and therefore fall under the classical explanations we have so far disregarded. 

There were additional assumptions however, that go into our conclusions. The first is that the liquids behaved as ideal free-electron systems. While for pure elements there is experimental evidence to suggest that the optical properties of liquid metals behave close to those of a Drude oscillator model, in even simple commercially available alloys like 304L stainless steel, this data is sorely lacking, thereby requiring us to transpose the free electron assumption from pure elements to alloys \cite{comins1972optical,miller1969optical}. The second main assumption is one of non-resonance. While electronic structure transitions are indicated in metals with complicated absorption spectra, in liquids, even in systems that nominally behave ideally, there may exist ordering and clustering that can be detected via electronic transport property measurements\cite{paras2024evidence}. Such clustering could produce structures that resonate and therefore alter how laser energy is coupled into a system, explaining why under laser illumination, non-equilibrium features emerge. These realities shall only be explored through experimental design that can enable us to access conditions of temperature and pressure relevant under the current processing conditions.


\section{Conclusions}
Sections of 304L plate were irradiated using Blue and IR frequencies. A fundamental change in the chemistry and morphology of the oxide layer was observed. This could not be explained using our conventional understanding of laser-matter interaction in the liquid metal. This data corroborates other evidence in the literature that suggests varying the laser frequency may be inducing as-yet determined chemical potential shifts in liquid metals.

\section{Acknowledgements}
The authors thank Levi Van Bastian for additive manufacturing laser tool operation.  This work was supported by the Sandia National Laboratory Directed Research and Development (LDRD) program. Sandia National Laboratories is a multimission laboratory managed and operated by National Technology and Engineering Solutions of Sandia, LLC., a wholly owned subsidiary of Honeywell International, Inc., for the U.S. Department of Energy’s National Nuclear Security Administration under Contract No. DE-NA0003525. This work describes objective technical results and analysis. Any subjective views or opinions that might be expressed in the paper do not necessarily represent the views of the U.S. Department of Energy or the U.S. Government. On behalf of all authors, the corresponding author states that there is no conflict of interest.

\bibliography{main}

\clearpage

\section{Appendix}

The figure below is a stylistic depiction of the ToF-SIMS data from Figures (\ref{TofSims}) and (\ref{TofSims2}).

\begin{figure}[h]
    
    \centering
    \includegraphics[width=.9\textwidth, height=.6\textheight]{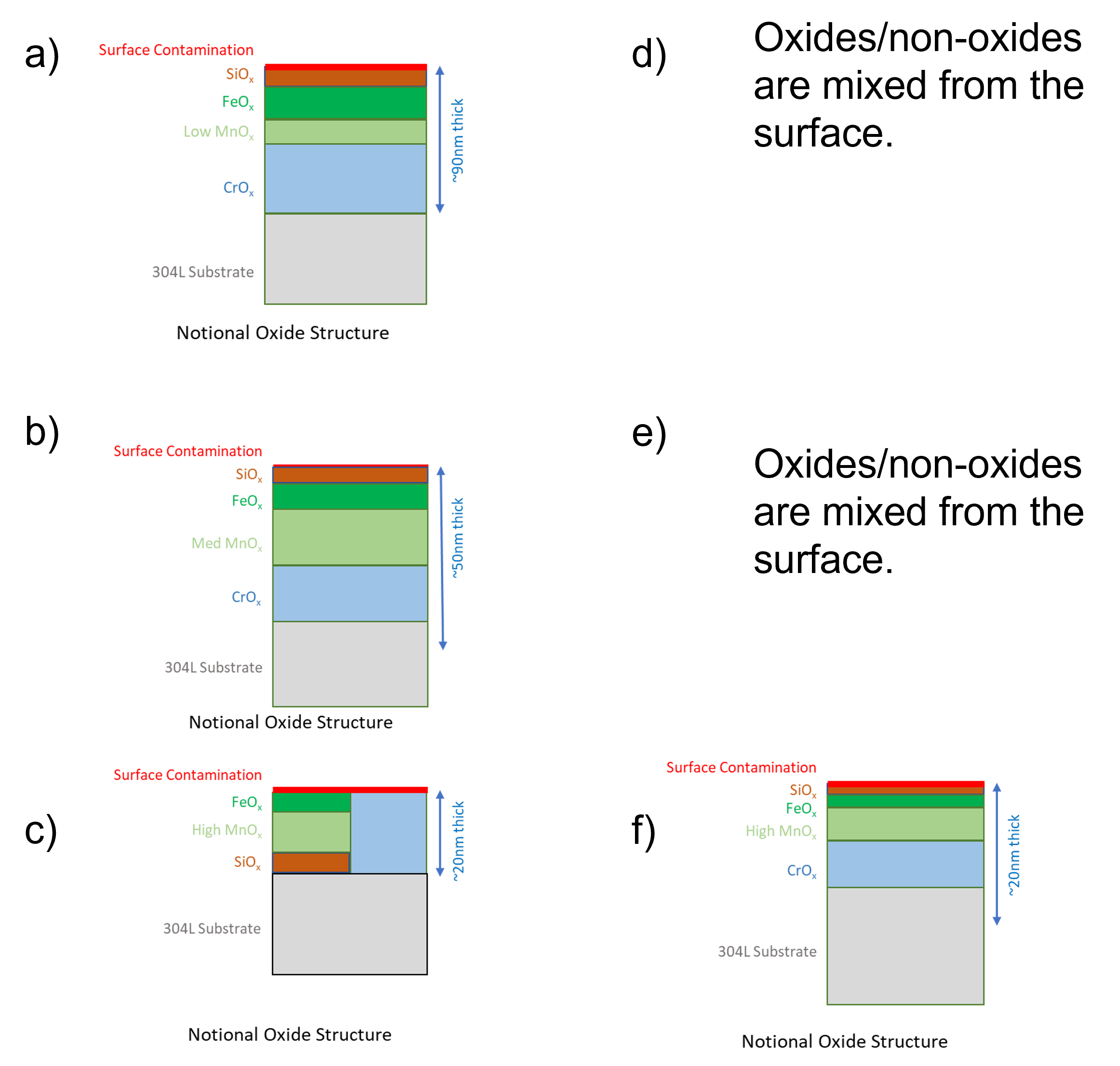}
  \label{appendixCartoon}
  \caption{This figure is a cartoon representation of the notional oxide layers based on the ToF SIMS data in Figure (\ref{TofSims}). The oxide layer gets thinner with increasing track speed in the case of the blue laser. Compared to IR, the oxide layer is thicker, with increased FeO$_{x}$ at the surface, compared to a Cr-rich layer in the IR that is thinner, and oxygen deficient.}
\end{figure}

\end{document}